# Investigation of magnetic characteristics of oxygen adsorbed $YbFe_2As_2$ single crystals

**S. Santhosh Raj[1], Nilotpal Ghosh[2], R. Navamathavan[1*]**

*[1]Division of Physics, School of Advanced Sciences, VIT University Chennai, Vandalur – Kelambakkam Road, Chennai – 600 127*
*[2]Science and Engineering Research Board, Department of Science and Technology, Vasant Square Mall, Vasant Kunj, New Delhi 110 070*

*Corresponding Author:
E-mail: n_mathavan@yahoo.com; navamathavan.r@vit.ac.in
Ph. No.: +91-44-3993 1258; FAX: +91-44-3993 2555

## ABSTRACT

Recent discovery of superconductivity in iron pnictides had attracted immense attention of the scientific community. The parent compounds were spin density wave (SDW) metals unlike the high-$T_C$ superconductors which were Mott insulators. In this present study, we synthesized single crystal of a new compound $YbFe_2As_2$ by using high temperature solution growth technique without flux. The $YbFe_2As_2$ single crystals had been systematically characterized by energy dispersive X-ray analysis (EDAX). The presence of oxygen was found by EDAX on the surfaces of grown $YbFe_2As_2$ single crystals which had been kept in air ambience for few months. The measurement of magnetization (M) versus temperature (T) using SQUID at constant magnetic field (H = 100 Oe) for $YbFe_2As_2$ had revealed an occurrence of sharp slope change around 140 K which was the known SDW transition temperature for the parent compound $BaFe_2As_2$. An additional slope change had been observed around 40 K. M vs T data at H= 10000 Oe had exhibited paramagnetic behavior. Result of M versus H measurements for $YbFe_2As_2$ at 2 K had shown that the saturation had not been achieved at H = 80000 Oe. We had carried out magnetization measurements for oxygen adsorbed $YbFe_2As_2$ ($YbFe_2As_2 : O_2$ ) and $BaFe_2As_2$ ($BaFe_2As_2 : O_2$) for comparative study also.

## 1. Introduction

There was considerable interest in the recently discovered iron-based superconductors because the superconducting transition temperature was as high as 56 K and superconductivity arose from antiferromagnetic or spin density wave (SDW) metallic parent compounds [1-4]. Furthermore, there were reports of the co-existence of superconductivity and magnetism particularly in the under doped region [5-7]. Parent compounds of iron-based superconductors was reported to undergo a tetragonal-to-orthorhombic structural transition upon cooling accompanied by the SDW transition and to show anomalous metallic behavior with indication of antiferromagnetic (AFM) ordering [8]. Strong influence of adsorbed oxygen (air) in polycrystalline $BaFe_2As_2$ to mask the spin state of Fe had been reported [9]. Recently, it was found that oxygen adsorbed $BaFe_2As_2$ ~~a~~ single crystal did not show SDW transition around 140 K in resistivity versus temperature measurement [10]. Hence, the iron-based oxy-arsenide superconductors provided a new platform to study the interplay between magnetism and superconductivity, as shown by many theoretical studies [11-13]. To our knowledge, there were no detailed reports available in the literature based on the magnetic properties of $YbFe_2As_2$ single crystals. Therefore, in this present investigation, we performed detailed experimental studies to understand the magnetic behavior of $YbFe_2As_2$ single crystals.

Herein, we reported the synthesis of $YbFe_2As_2$ single crystals by using high temperature solution growth technique without flux. The grown single crystals had been systematically characterized by energy dispersive X-ray analysis (EDAX) and scanning electron microscopy (SEM). The magnetic characteristics of $YbFe_2As_2$ single crystals were performed by using SQUID. For comparative analysis, we had performed magnetic characteristic of oxygen adsorbed in $YbFe_2As_2$ and $BaFe_2As_2$ single crystals and the results were reported in detail.

## 2. Experimental methods

$YbFe_2As_2$ were synthesized by high temperature solution growth technique without using flux [14]. Elemental Yb, Fe, As were added and the reaction product was homogenized in an agate mortar and placed in a alumina crucible which was vacuum sealed in a quartz ampoule. The sealed ampoule was placed in a programmable furnace and heated to 1300 $^{o}$C. Then it was cooled at 3 $^{o}$C per hour up to 95$^{o}$ C and again cooled at 50 $^{o}$C per hour up to room temperature. $BaFe_2As_2$ had been grown by high temperature solution growth technique using FeAs as self-flux [15]. The Elemental Ba, Fe, and As were added to FeAs in the ratio of $BaFe_2As_2$ : FeAs = 1:2 and placed in a programmable furnace and heated to 1100$^{o}$ C. Then it was cooled at 75 hours up to 900$^{o}$ C and in 4 hours to room temperature. Finally, the synthesized compound was taken out from the ampoule. Fig. 1(a) showed the temperature controlling program during the synthesis of $YbFe_2As_2$ single crystals. Fig. 1(b) showed the photograph of as-grown $YbFe_2As_2$ single crystals.

Both the crystals were kept in air ambience for several months and composition of the single crystals had been determined from energy dispersive X-ray analysis (EDAX) analysis. We carried out the magnetization measurements of $YbFe_2As_2$ and $BaFe_2As_2$ crystals by using SQUID.

## 3. Results and Discussion

In order to understand the chemical composition present in the as-grown $YbFe_2As_2$ single crystals, we performed the EDAX measurement. Fig. 2 showed the EDAX data of $YbFe_2As_2$:$O_2$ single crystal. Prominent signals from the Yb, Fe, and As elements were clearly observed from the EDAX data. In addition, as could be seen, the intensity peak for oxygen was also observed. The oxygen was adsorbed on the surface of as grown $YbFe_2As_2$ crystals due to exposure in air for several months. Adsorbed $O_2$ was known to be an effective electron acceptor forming in $O_{2-}$ which could permeate the $YbFe_2As_2$ single crystal and lead to the formation of spin clusters around Fe.

Oxygen adsorption on crystal surfaces was accompanied by an increase in the work function. Oxygen adsorption might cause some surface reconstruction and it could penetrate beneath the surface layer [16]. As a result of that Fe spin state got affected in the $YbFe_2As_2$ crystal [9].

The magnetic properties of as-prepared $YbFe_2As_2 : O_2$ single crystals were analyzed by means of SQUID measurements. Here, we also measured the magnetic properties for $BaFe_2As_2 : O_2$ single crystals for the comparative analysis. Figs. 3(a) and 3(b) showed the measurements of magnetization (M) as a function of temperature (T) at constant H= 100 Oe for $YbFe_2As_2 : O_2$ and $BaFe_2As_2 : O_2$ crystals, respectively. A sharp peak was observed in magnetization data around 140 K which had generally been attributed to the structural transition from tetragonal to orthorhombic accompanying SDW transition for the well-studied parent compound of $BaFe_2As_2$ [8]. An additional slope change had been observed around 40 K. In case of oxygen adsorbed $BaFe_2As_2$ crystals, magnetization measurement did not show any peak around 140 K, but showed sharp slope change around 40 K.

$YbFe_2As_2 : O_2$ crystals showed significant drop in magnetization around 140 K and later around 40 K, which could be indication for development of anti-ferromagnetic order Similar drop in magnetization was observed in $BaFe_2As_2 : O_2$ around 40 K which could be attributed to antiferromagnetic order also. This kind of effects might be prominent at small applied field [17].

Fig. 4 showed the temperature dependence of magnetization (M) of $YbFe_2As_2 : O_2$ and $BaFe_2As_2 : O_2$ single crystals under a high magnetic field of H = 10000 Oe. There was no peak noticed in magnetization around 140 K and 40 K for both type of the crystals at H = 10000 Oe. It was observed that the magnetization slowly decreased with increasing temperature suggesting paramagnetic behaviour. The magnetization remained almost constant and slightly varying with temperature above 140 K which could be an indication of Pauli-paramagnetism [8,17].

Fig. 5 showed the measurements of magnetization (M) as function of magnetic field (H) at T = 2 K and 300 K for $YbFe_2As_2$ : $O_2$ crystals. The saturation of moment had not been achieved for $YbFe_2As_2$ : $O_2$ sample at T = 2 K by applying field up to H = 80000 Oe. This was probably due to Yb moments present in the $YbFe_2As_2$ crystals.

Fig. 6 showed the results of M versus H measurements at T = 2 K for both $YbFe_2As_2$ : $O_2$ and $BaFe_2As_2$ : $O_2$ crystals. It can be observed that the saturation was achieved for $BaFe_2As_2$ : $O_2$ crystal at H = 80000 Oe. On the other hand, the magnetization of $YbFe_2As_2$ : $O_2$ single crystals monotonically increased with increasing magnetic field.

## 4. Conclusions

Single crystals of $YbFe_2As_2$ had been grown by high temperature solution growth technique and characterized. Oxygen adsorbed $YbFe_2As_2$ showed sharp peak around 140 K and also additional slope change had been observed around 40 K in magnetization measurement at H = 100 Oe. But oxygen adsorbed $BaFe_2As_2$ crystals did not show any peak in similar magnetization measurement around 140 K which was well known SDW transition temperature for the parent compound $BaFe_2As_2$. However, $BaFe_2As_2$ : $O_2$ showed sharper slope change in magnetization compared to what was shown by $YbFe_2As_2$ : $O_2$ around 40 K. The magnetization measurement at H = 10000 Oe showed that $YbFe_2As_2$:$O_2$ and $BaFe_2As_2$ : $O_2$ were paramagnetic in nature. Since, Yb had a moment, the saturation in M versus H measurements had not been achieved for $YbFe_2As_2$ at H = 80000 Oe at T = 2 K. The role of Yb moments was not yet understood.

## Acknowledgments

NG and SR wanted to thank DST-SERB, Government of India under project SR/S2/CMP-0123/2012 for financial support. NG wanted to thank UGC-DAE Consortium for Scientific Research, Indore for providing SQUID facility for magnetic measurements.

**References**

[1] Y. Kamihara, T. Watanabe, M. Hirano, H.J. Hosono, J. Am. Chem. Soc., 130 (2008) 3296.

[2] K. Ishida, Y. Nakai, H.J. Hosono, J. Phys. Soc. Jpn., 78 (2009) 062001.

[3] D.C. Johnston, Adv. Phys., 59 (2010) 803.

[4] J. Paglione, R.L. Greene, Nature Phys., 6 (2010) 645

[5] M.D. Lumsden, A.D. Christianson, J. Phys.: Condens. Matter., 22 (2010) 203203;

[6] Y.J. Uemura, Nature Mater., 8 (2009) 253.

[7] S.C. Speller, T.B. Britton, G. Hughes, S. Lozano Perez, A.T. Boothroyd, E. Pomjakushina, K. Conder, C.R.M. Grovenor, Appl. Phys. Lett., 99 (2011) 192504.

[8] M. Rotter, M. Tegel, D. Johrendt, I. Schellenberg, W. Hermes, H. Pottgen, Phys. Rev. B, 78 (2008) 020503(R).

[9] O. Khasanov, U. Reichel, E. Dvilis, A. Khasanov, J. Phys.: Condens. Matter., 23 (2011) 3422.

[10] Nilotpal Ghosh, S. SanthoshRaj, AIP Conference Proceedings, 1665 (2015) 100009.

[11] K. Haule, J.H. Shim, G. Kotliar, Phys. Rev. Lett., 100 (2008) 226402.

[12] C. Cao, P.J. Hirschfeld, H.P. Cheng, Phys. Rev. B, 77 (2008) 220506(R).

[13] D.J. Singh, M.H. Du, Phys. Rev. Lett., 100 (2008) 23700.

[14] H.S. Jeevan, Deepa Kasinathan, Helge Rosner, Philipp Gegenwart, Phys. Rev. B, 83 (2011) 054511.

[15] A. Mani, N. Ghosh, S. Paulraj, A. Bharathi, C.S. Sundar, Solid State Communications, 150 (2010) 1940.

[16] V.F. Kiselev, O.V. Krylov, Adsorption and Catalysis on Transition Metals and Their Oxides, Springer-Verlag, Berlin, (1989).

[17] C. Krellner, N. Caroca-Canales, A. Jesche, H. Rosner, A. Ormeci, C. Geibei, Phys. Rev. B, 78 (2008) 100504 (R).

## Figures Caption

**Fig. 1.** (a) Temperature controlling program used for the growth of $YbFe_2As_2$ and (b) $YbFe_2As_2$: $O_2$ single crystals.

**Fig.2 .** The result of EDAX experiments on $YbFe_2As_2$ : $O_2$ single crystals after few days of growth.

**Fig. 3.** The plot of magnetization as function of temperature at H= 100 Oe for (a) $YbFe_2As_2$: $O_2$ and (b) $BaFe_2As_2$ : $O_2$ crystals.

**Fig. 4.** Magnetization measurements of $YbFe_2As_2$ : $O_2$ and $BaFe_2As_2$ : $O_2$ crystals.

**Fig. 5.** Results of M versus H measurements of at T = 2 K and 300 K for $YbFe_2As_2$ : $O_2$ crystals.

**Fig. 6.** Results of M versus H measurements of at T = 2 K for $YbFe_2As_2$ : $O_2$ and $BaFe_2As_2$ : $O_2$ crystals.

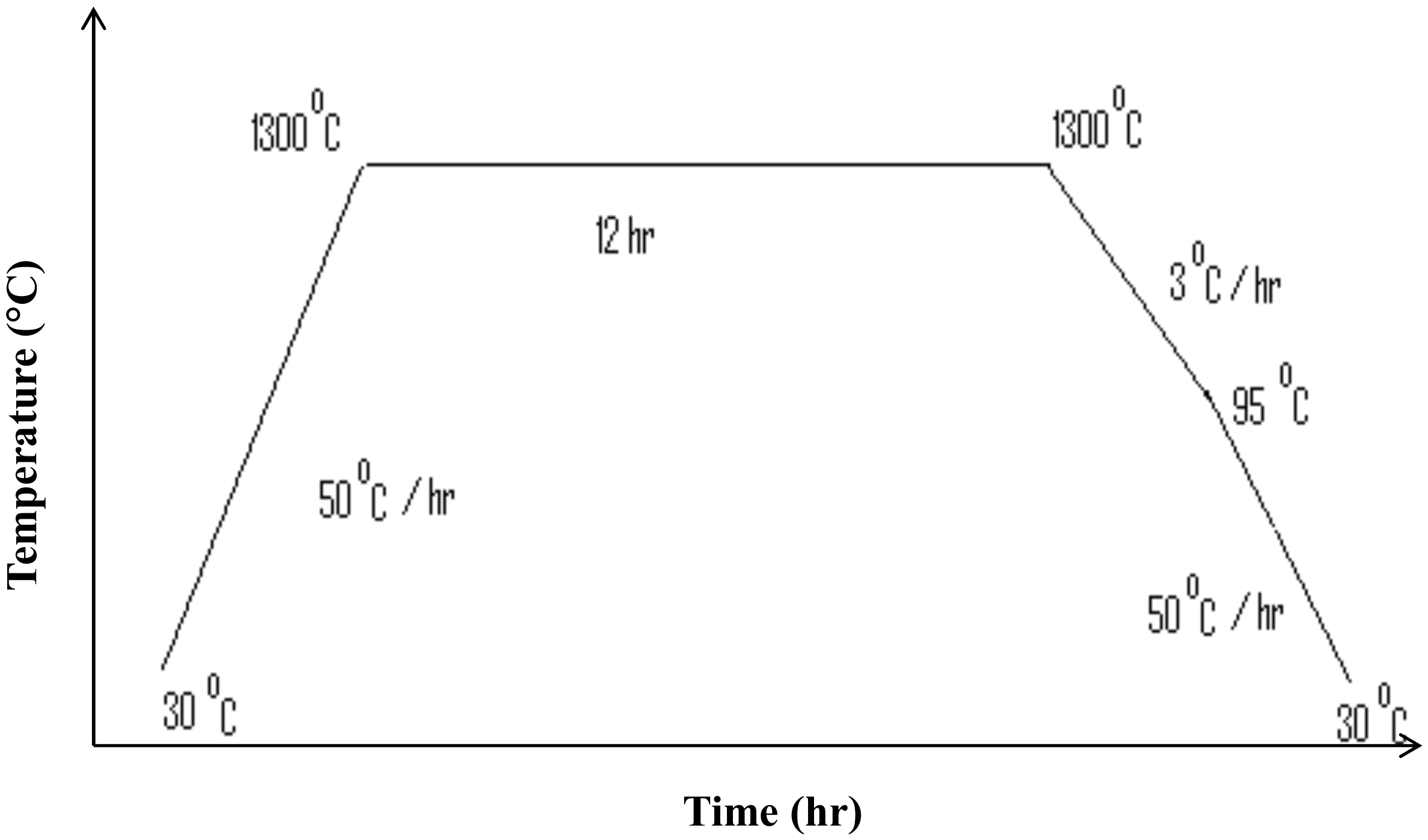

1300 °C
1300 °C
12 hr
3 °C / hr
95 °C
50 °C / hr
50 °C / hr
30 °C
30 °C
Temperature (°C)
Time (hr)

**Fig. 1(a).**

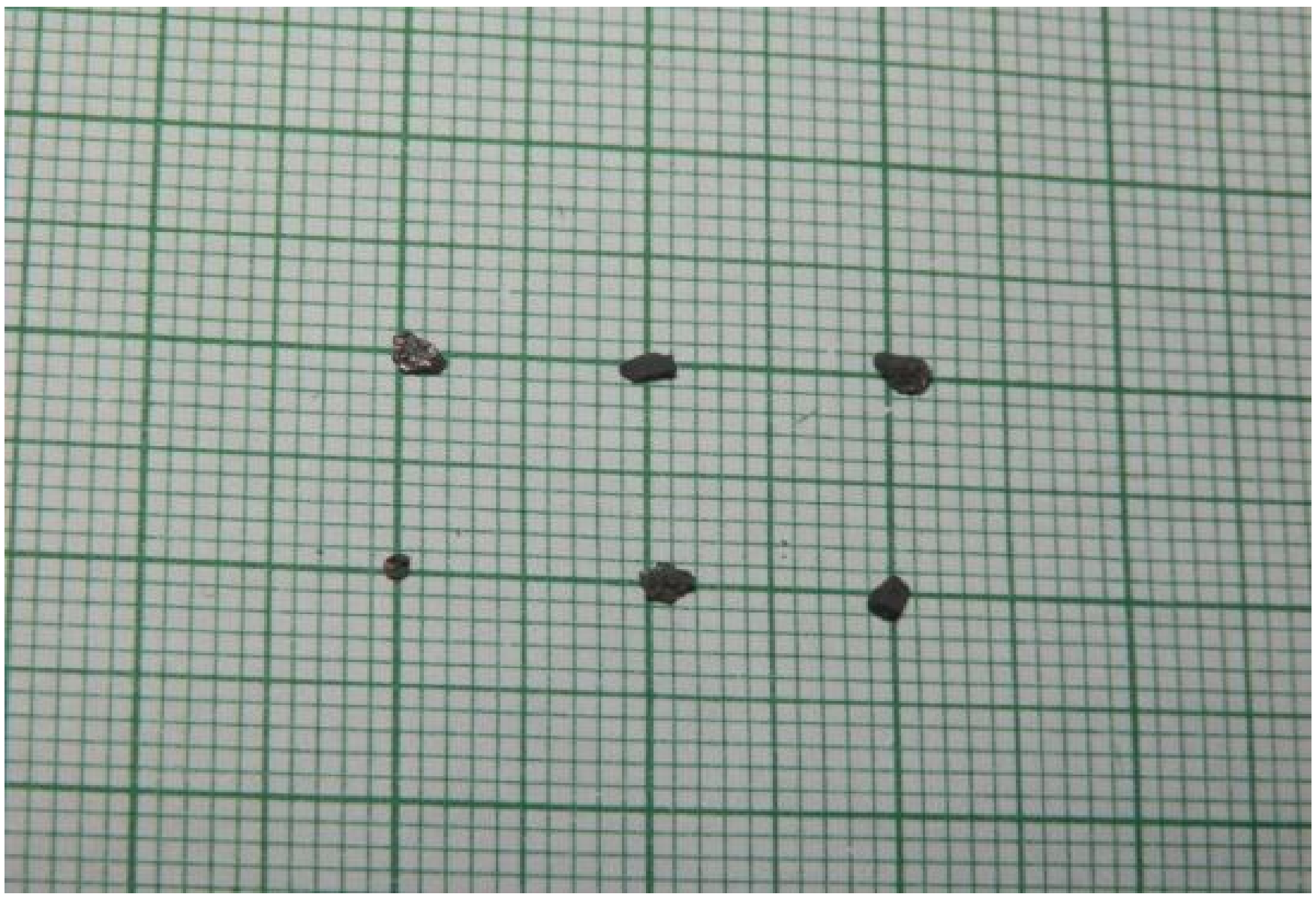

**Fig. 1(b).**

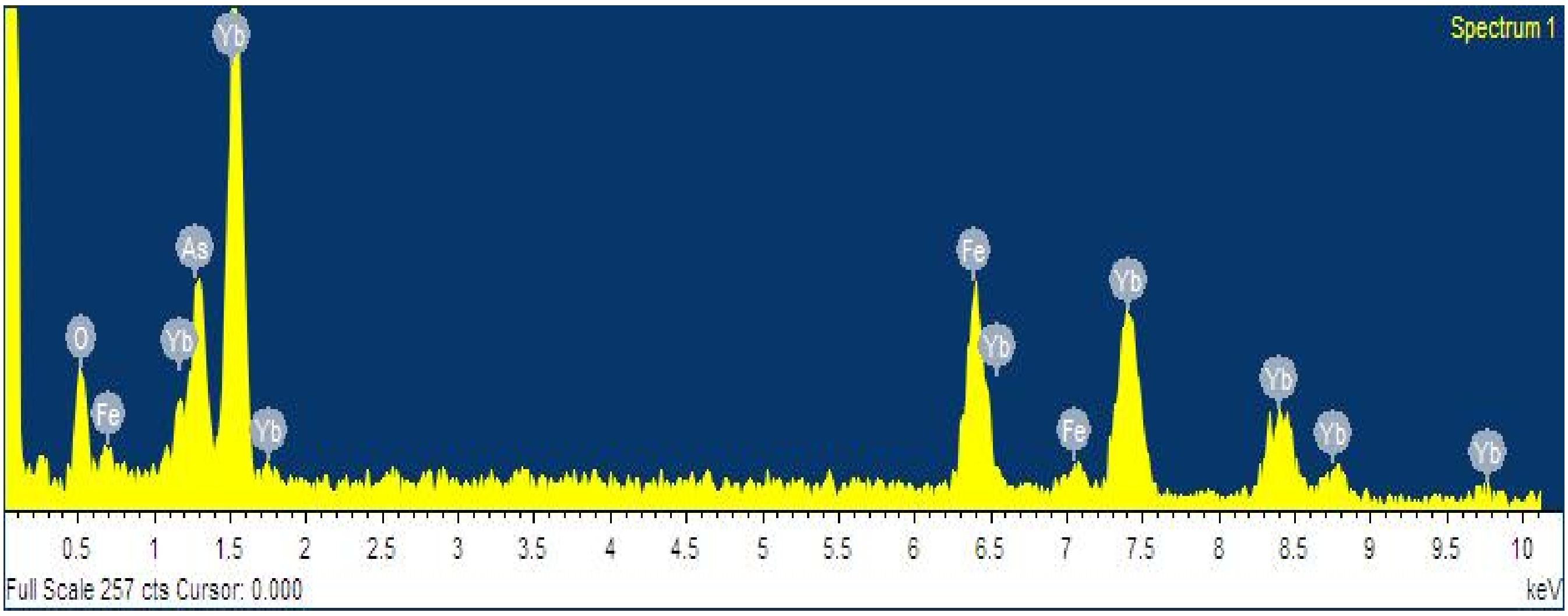
Spectrum 1
Yb
As
O
Yb
Fe
Yb
Fe
Yb
Yb
Fe
Yb
Yb
Yb
0.5 1 1.5 2 2.5 3 3.5 4 4.5 5 5.5 6 6.5 7 7.5 8 8.5 9 9.5 10
Full Scale 257 cts Cursor: 0.000
keV

**Fig. 2.**

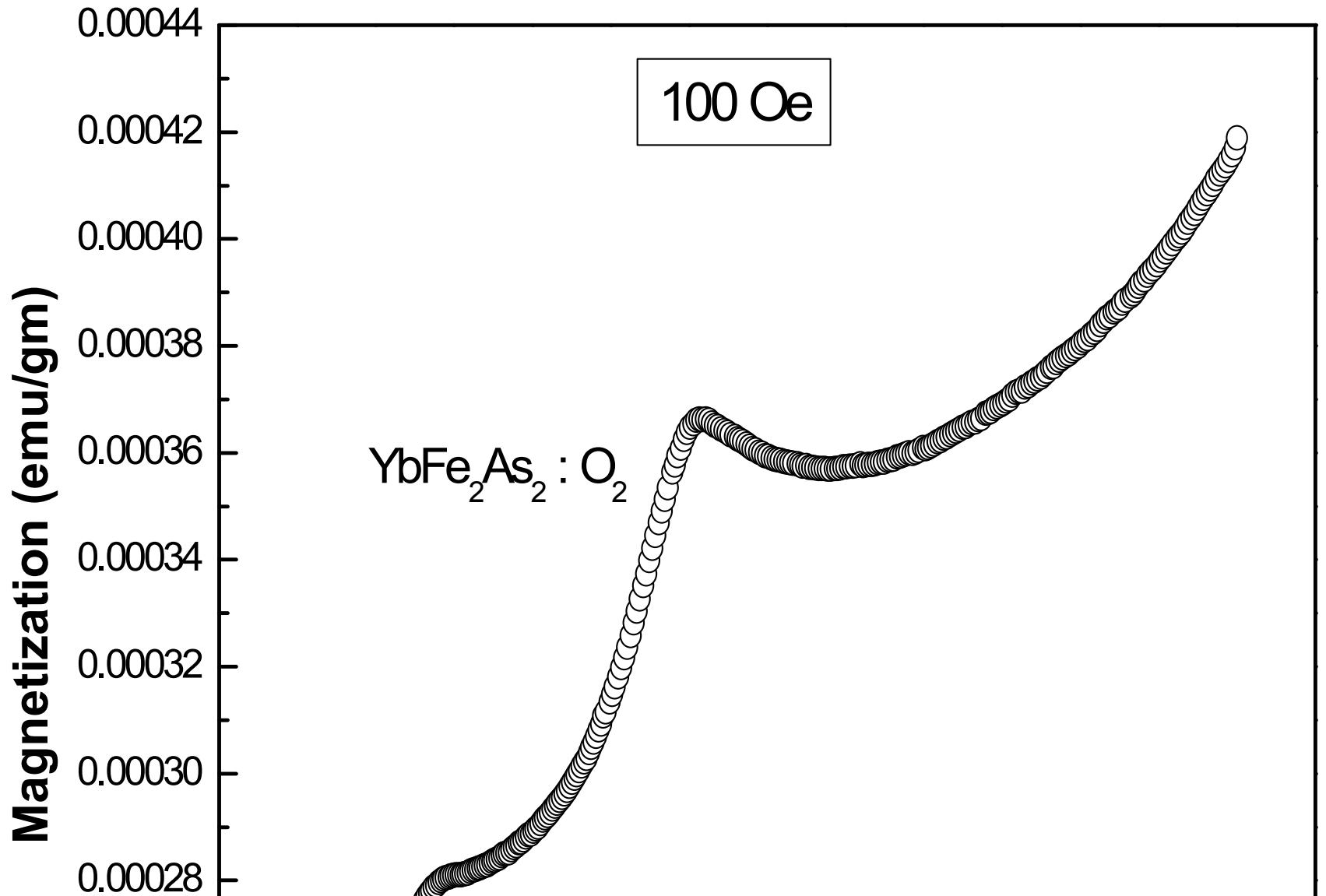

**Fig. 3(a).**

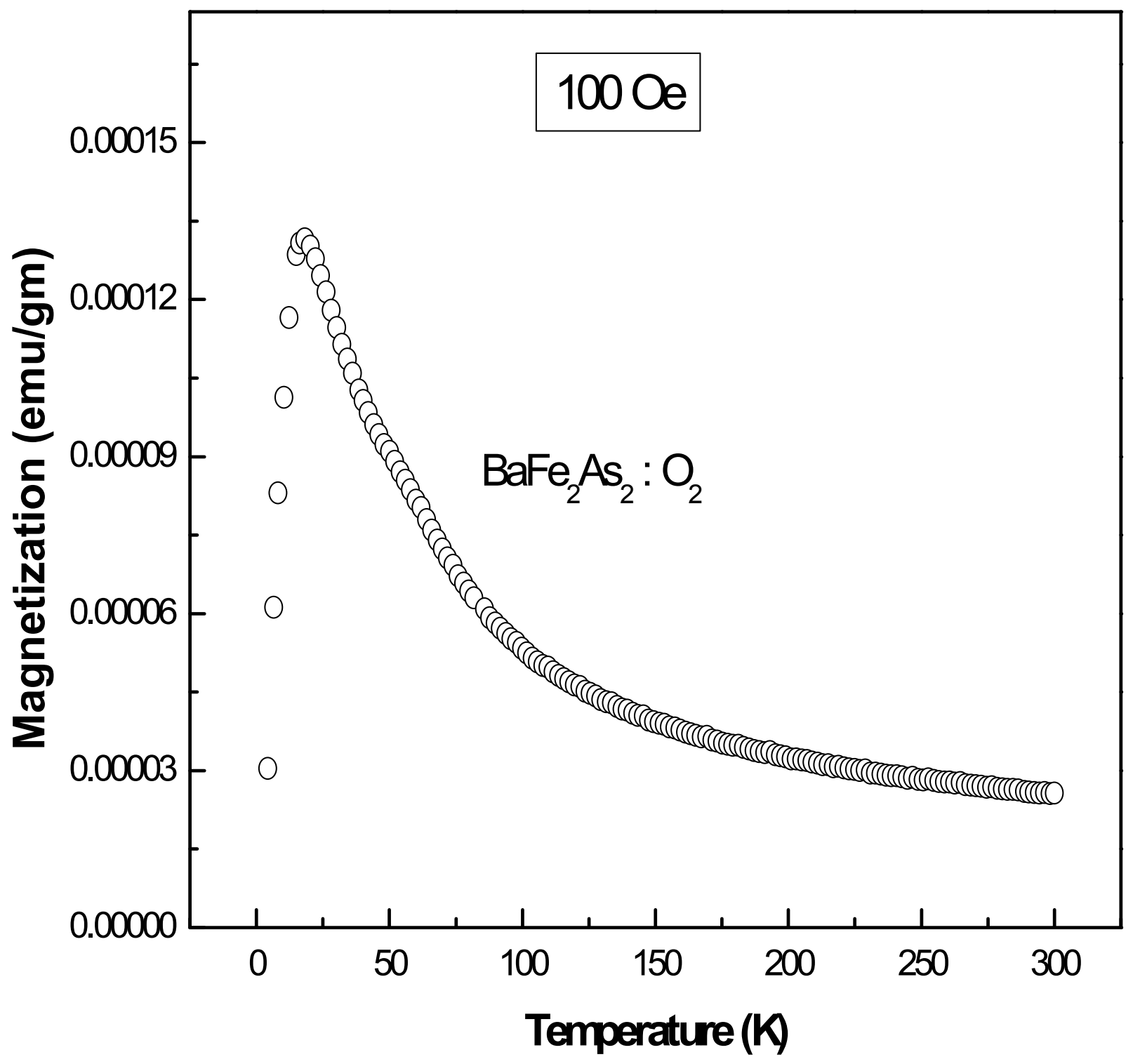

**Fig. 3(b).**

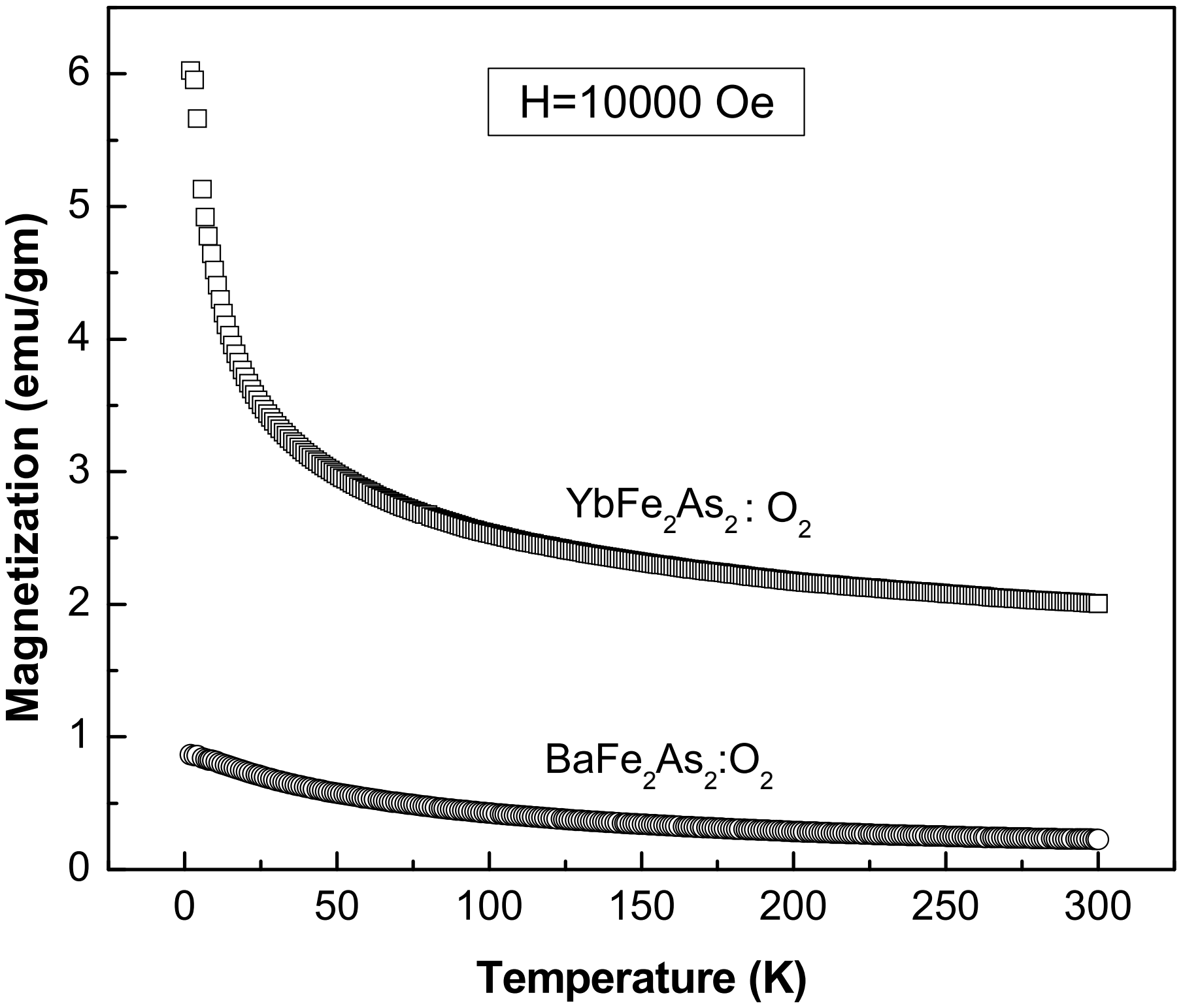

**Fig. 4.**

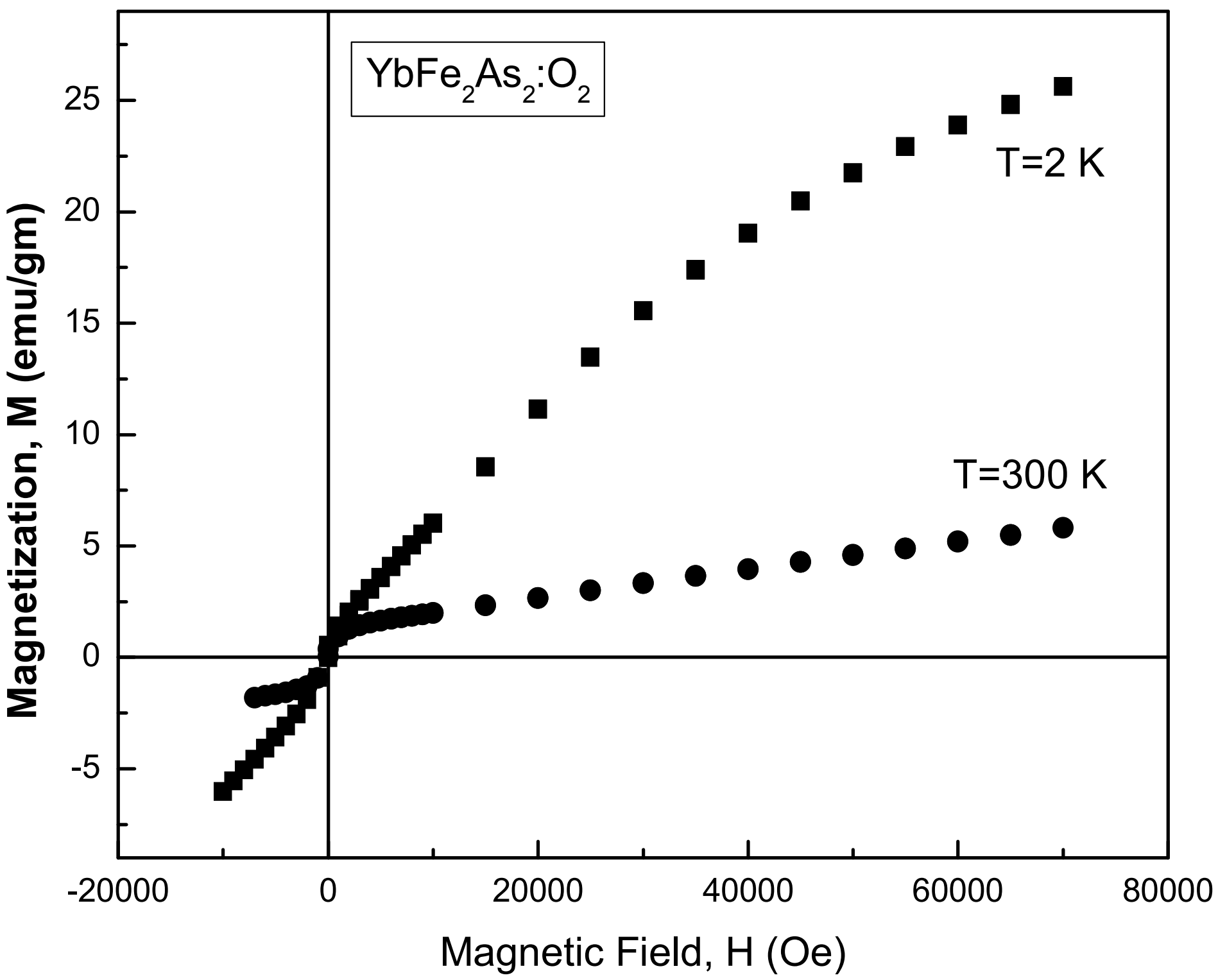

**Fig. 5.**

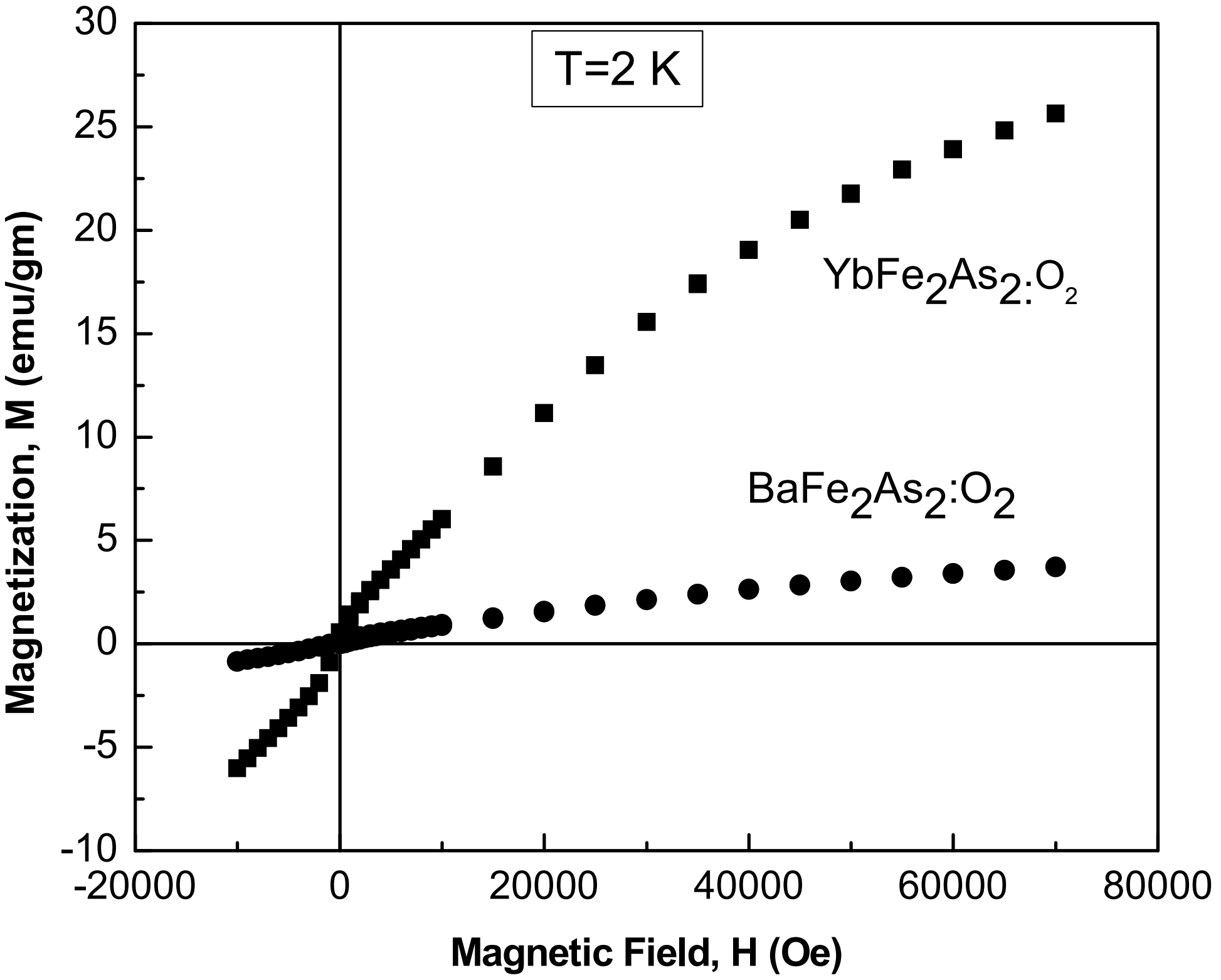

**Fig. 6.**